\documentclass[
	a4paper, % Paper size, use either a4paper or letterpaper
	10pt, % Default font size, can also use 11pt or 12pt, although this is not recommended
	unnumberedsections, % Comment to enable section numbering
	twoside, % Two side traditional mode where headers and footers change between odd and even pages, comment this option to make them fixed
]{LTJournalArticle}

\runninghead{} % A shortened article title to appear in the running head, leave this command empty for no running head

\footertext{\textit{RISC-V Summit Europe, Munich, 24-28th June 2024}} % Text to appear in the footer, leave this command empty for no footer text

\title{RISC-V for HPC: Where we are and where we need to go} % Article title, use manual lines breaks (\\) to beautify the layout

\author{
	Nick Brown\textsuperscript{1}\thanks{Corresponding author: \href{mailto:n.brown@epcc.ed.ac.uk}{\tt n.brown@epcc.ed.ac.uk}}
}

\date{\footnotesize\textsuperscript{\textbf{1}}EPCC, University of Edinburgh, Bayes Centre, 47 Potterrow, Edinburgh, United Kingdom}

\renewcommand{\maketitlehookd}{%
	\begin{abstract}
Funded by the UK ExCALIBUR H\&ES exascale programme, since early 2022 we have provided a RISC-V testbed for HPC to offer free access for scientific software developers to experiment with RISC-V for their workloads. Based upon our experiences of providing access to RISC-V for the HPC community, and our involvement with the RISC-V community at large, in this extended abstract we summarise the current state of RISC-V for HPC and consider the high priority areas that should be addressed to help drive adoption. 
	\end{abstract}
}

\begin{document}

\maketitle % Output the title section

%----------------------------------------------------------------------------------------
%	ARTICLE CONTENTS
%----------------------------------------------------------------------------------------

\section{Introduction}

Whilst RISC-V has grown rapidly in areas such as embedded computing, it is yet to gain significant traction in High Performance Computing (HPC). However, as we move further into the exascale era the HPC community will be faced by a range of new challenges, for instance to decarbonise their workloads, and there is the potential for RISC-V to play an important role.

We stood up the ExCALIBUR H\&ES RISC-V testbed to provide the ability for the HPC community to experiment with RISC-V for their workloads. Offering free access to RISC-V for developers, in the two years that the testbed has been running we have seen a phenomenal rate of change around aspects of RISC-V related to HPC, with interest by the HPC community also growing steadily. There are however still numerous activities that must be addressed to help drive adoption and make RISC-V an attractive proposition for HPC.

\section{RISC-V \& HPC: Where we are}
\subsection{Hardware}
When we first stood up our RISC-V HPC testbed in early 2022, it was a very significant challenge to gain access to physical RISC-V hardware. The HiFive Unmatched with its four cores and 16GB of DRAM, which the Monte Cimone Cluster was built around, had recently been discontinued. Initially, low powered single-core boards, such as the MangoPi MQ-Pro, started to appear. Aimed very much at the embedded computing market, they were rather lacklustre for any but the simplest workloads and also required significant reconfiguration from the stock OS image, requiring rebuilding the bootloader and Linux kernel for instance, to integrate into our testbed.

Then, in mid 2022, the dual core VisionFive V1 which contains 8GB became available. This was a considerable improvement, and was then followed towards the end of 2022 by the quad core VisionFive V2. However, all the machines mentioned so far are Single Board Computers (SBCs), and whilst they are interesting to experiment with, it was still not possible to obtain commodity RISC-V hardware that could be considered any sort of serious proposition for HPC.

This changed in summer 2023, when Sophon announced the 64-core SG2042 which is designed for high performance workloads. Given the large number of cores coupled with up to 128GB of memory on the Milk-V Pioneer workstation, this is for the first time a serious RISC-V option for HPC workloads. Table \ref{tab:npb_perf} reports a performance comparison of the SG2042 against the RISC-V SBCs mentioned in this section. Running kernels from NASA's NAS Parallel Benchmark (NPB) suite, which broadly represent key algorithmic patterns to be found in Computational Fluid Dynamics (CFD) codes, over all cores of the CPU and parallelised via OpenMP, Table \ref{tab:npb_perf} reports, for each SBC, the number of times slower that SBC is than the SG2042.

\begin{table}[]
    \centering
    \caption{Number of times slower each RISC-V SBC is than the SG2042, running NPB kernels parallelised with OpenMP over all cores (SBC number of cores in brackets)}
    \label{tab:npb_perf}
    \begin{tabular}{|c|c|c|c|c|}
    \hline
      \textbf{Benchmark} & V2 (4) & V1 (2) & SiFive (4) & MangoPi (1) \\
     \hline
	  IS & 17.5 & 132.7 & 54.1 & 105.3\\        
	  MG & 26.7 & 138.5 & 52.6 & 322.9 \\
        EP & 24.7 & 81.6 & 41.7 & 125.6 \\
        CG & 17.3 & 65.6 & 105.3 & 215.9 \\
        FT & 13.3 & 58.5 & 32.3 & DNR\\
    \hline
    \end{tabular}
\end{table}

It can be seen from Table \ref{tab:npb_perf} that the SG2042 delivers very considerably higher performance than the previous commodity available RISC-V hardware, however benchmarking \cite{brown2023risc} has shown that it is still around between four and eight times slower than modern x86-based hardware commonly used in HPC.

\subsection{Software}
The RISC-V software ecosystem has matured rapidly, and we have found that the vast majority of our user's HPC codes build out of the box, albeit with some changed required to the build system especially when cross compiling for RISC-V. Indeed, the RISC-V interactive landscape resource \cite{software-landscape} is somewhat pessimistic as many other important HPC libraries that are not listed, such as PETSc, NetCDF and HDF5, also build and run on RISC-V (albeit unoptimised). HPC software infrastructure support is also present, and the fact that the Slurm job submission system is available on RISC-V, as well as other miscellaneous functionality such as networked file systems, means that a RISC-V system can be set up matching the configuration of main stream HPC machines and feel familiar to users.

\subsection{Community}
There is a growing interest in RISC-V by the HPC community, thanks in large part due to the efforts of the RISC-V HPC SIG. Workshops organised at key HPC conferences, such as ISC and SC, have resulted in a large audience with a wide range of topics being covered. Indeed, as an example, at the SC23 RISC-V workshop the HPC community's popular publication, HPCWire, wrote an in depth news article around RISC-V and HPC based on details shared by vendors and the keynote speaker, the CTO of RISC-V International, during the session.

\section{RISC-V \& HPC: What is missing}
Undoubtedly, more powerful RISC-V commodity available hardware to close the performance gap against the x86 CPUs currently employed for HPC would be most welcome. Indeed, there are several promising activities such as the SG2044 CPU from Sophon which will provide RVV v1.0 and increased memory performance over the SG2042. Moreover, RISC-V based accelerators such as Esperanto's ET-SoC, InspireSemi's Thunderbird, and Tenstorrent's Greyskull and Wormhole have great potential. Indeed, the ET-SoC-1 with 1000 RISC-V cores is already available, and the Greyskull and Wormhole cards available to buy from the Tenstorrent website.

A common question people tend to ask is \emph{why RISC-V for HPC?} Specifically, what are the key benefits that RISC-V can provide over and above other architectures for high performance workloads, and ultimately why would an HPC centre invest in RISC-V? It is our view that, to drive RISC-V adoption in HPC, it is critical that we can convincingly answer this question and provide evidence around where RISC-V can be beneficial. It is our opinion that RISC-V is most likely to initially make inroads in the accelerator space, where HPC centres purchase a number of RISC-V accelerator cards and fit these to existing machines. The ability to specialise the architecture to suit a specific class of problem could potentially deliver considerable energy efficiency improvements compared to, for instance, GPUs. Energy efficiency is becoming increasingly important as supercomputing operators are looking to decarbonise. It is our view that RISC-V based accelerators might also help drive adoption of RISC-V based CPUs due to the benefits of having a unified ISA across the CPU and accelerator. Indeed, in recent years we have seen closer integration between GPUs and CPUs in HPC by the provision of a unified memory space, with obvious benefits, and RISC-V provides the potential to push this a step further by unifying the ISA and programming model.

\subsection{Software ecosystem}
There is considerable work going on in and around the RISC-V software ecosystem and initiatives such as RISE have made a very positive impact on the overarching maturity. It is important that HPC connects with these activities and that we as the community develop a prioritised list of HPC libraries and applications that need to be ported to RISC-V and optimised for the architecture.

An especially important area to consider for HPC development is the provision of mature performance profiling tools. Whilst \emph{perf} is supported on RISC-V, and using it it is possible to track the CPU hardware counters during code execution, RISC-V currently lacks a high level profiling tool such as CrayPat, Vtune, or Forge that HPC developers often rely upon.

\section{Conclusions}
The advances in the RISC-V ecosystem have been phenomenal since we began our HPC testbed two years ago, and with a range of new hardware anticipated for 2024 it looks highly likely that this pace will continue to accelerate. It is important that RISC-V and the HPC community continue to work together, identify the key software building blocks required and are able to make a strong case for the role of RISC-V in HPC.

%\section{Acknowledgement}
%The authors would like to thank the ExCALIBUR H\&ES RISC-V testbed for access to compute resource and for funding this work. Interested readers can access the testbed at \emph{https://riscv.epcc.ed.ac.uk}. For the purpose of open access, the author has applied a Creative Commons Attribution (CC BY) licence to any Author Accepted Manuscript version arising from this submission.

%Furthermore, a repository of know-how around rebuilding the kernels (and different kernels) for the hardware currently available would also be advantageous.

% Not just CPU but also accelerators and specialisation - the argument needs to be made however!

%----------------------------------------------------------------------------------------
%	 REFERENCES
%----------------------------------------------------------------------------------------

\printbibliography % Output the bibliography

%----------------------------------------------------------------------------------------

\end{document}